\documentclass{kluwer}    

\newcommand{\xmmn}{\mbox{\em XMM-Newton}}

\newcommand{\exosat}{\mbox{\em EXOSAT}}

\newcommand{\etal}{\mbox{et\ al.\ }}

\newcommand{\msun}{\,\mbox{$\mbox{M}_{\odot}$}}

\newcommand{\aanda}  {A\&A\nolinebreak }

\newcommand{\mn}     {MNRAS\nolinebreak }
\newdisplay{guess}{Conjecture}

\begin{document}                                                                                   
\begin{article}
\begin{opening}         
\title{The Seyfert-LINER Galaxy NGC 7213: An \emph{XMM-Newton} Observation\thanks{This work is based on observations obtained with \xmmn, an ESA science
 mission with instruments and contributions directly funded by ESA Member
 States and the USA (NASA).}} 
\author{R.~L.~C. \surname{Starling}$^{1,2}$, M.~J. \surname{Page}$^{2}$, G. \surname{Branduardi-Raymont}$^{2}$, A.~A. \surname{Breeveld}$^{2}$, R. \surname{Soria}$^{2}$ and K. \surname{Wu}$^{2}$}  
\runningauthor{R.~L.~C. Starling and M.~J. Page et al.}
\runningtitle{The Seyfert-LINER galaxy NGC 7213: an \emph{XMM-Newton} observation}
\institute{$^{1}$Astronomical Institute, University of Amsterdam \\$^{2}$Mullard Space Science Laboratory, University College London}

\begin{abstract}
We examine the \emph{XMM} X-ray spectrum of the LINER-AGN NGC 7213, which is best fit with a power law, K$\alpha$
emission lines from Fe I, Fe XXV and Fe XXVI and a soft X-ray collisionally ionised thermal plasma with $kT = 0.18^{+0.03}_{-0.01}$ keV. We find a luminosity of 7$\times 10^{-4} L_{\rm Edd}$, and a lack of soft X-ray excess emission, suggesting a truncated accretion disc. NGC~7213 has intermediate 
X-ray spectral properties, between those of the weak AGN found
in the LINER M\,81 and higher luminosity Seyfert galaxies.
This supports the notion of a continuous sequence of X-ray properties from the Galactic Centre through LINER galaxies to Seyferts, likely 
determined by the amount of material available for accretion
in the central regions.
\end{abstract}
\keywords{X-rays: galaxies - galaxies: active - galaxies: Seyfert - galaxies: individual:
NGC 7213}
\end{opening}           

\section{Introduction}  
Low-ionisation nuclear emission-line region (LINER) galaxies are  
characterised by optical emission-line ratios which indicate
a low level of ionisation \cite{Heckman}. 
The origin of these emission lines is still
the subject of debate: the lines are attributed either to shock heating (Baldwin, Phillips and Terlevich, 1981) or to photoionisation
by a central AGN (Ferland and Netzer, 1983; Halpern and Steiner, 1983). 
NGC 7213 is a nearby ($z$=0.006) S0 galaxy with AGN and LINER 
characteristics. It is clear that there is an AGN in this source, classified as a Seyfert 1 \cite{phillips}.  
Optical emission lines are observed from this galaxy with
velocities ranging from 200 to 2000 km s$^{-1}$ FWHM (Filippenko and Halpern
1984, hereafter FH84). FH84 argue that this emission comes from photoionisation by the AGN of clouds spanning a range of densities and velocities.

Since its discovery as a low luminosity X-ray source \cite{HEAO2}
NGC 7213 has
been observed 
with several X-ray missions, showing a power law shaped spectrum with an Fe I K$\alpha$ line. Excess emission has been detected at higher energies, best explained as weak narrow emission lines from highly ionised iron. The same data have no significant reflection hump, suggesting that the Fe I K$\alpha$ line originates in Compton-thin material (Bianchi et al., 2003, hereafter B03, \emph{BeppoSAX} PDS+\emph{XMM} pn).
The presence of a soft X-ray excess in NGC~7213 was also inferred by
those observations and \exosat\ results \cite{exosat}.

Here we present the full \emph{XMM} observation of NGC~7213 including high resolution spectra from the RGS instruments. Identifying the physical mechanisms producing the X-ray
emission may help to reveal the origin of the optical emission lines
where at present neither shock heating nor photoionisation by the AGN can be
ruled out. We also discuss the relationship between Seyfert galaxies and LINERs. 
\section{\emph{XMM} observations and spectral fitting}
NGC 7213 was observed on 2001 May 28/29 with \emph{XMM} in the RGS GT Programme. The exposure times are 46448s for MOS1, 42201s for pn and 
46716s for each RGS instrument. The source is piled-up in MOS2. The MOS1 and pn spectra 
were combined using the method of
Page, Davis and Salvi (2003). The data were processed with \emph{XMM} SAS versions 5.2 and 5.4, and analysed using XSPEC v11.2.
The Galactic column used in all fits is $N_{\rm H}$=2.04$\times$10$^{20}$ cm$^{-2}$ \cite{gal}, errors are 90\% confidence for 1 interesting parameter and line energies are rest frame values.
\begin{figure}
\centering
\vspace*{7cm}
\leavevmode
\includegraphics{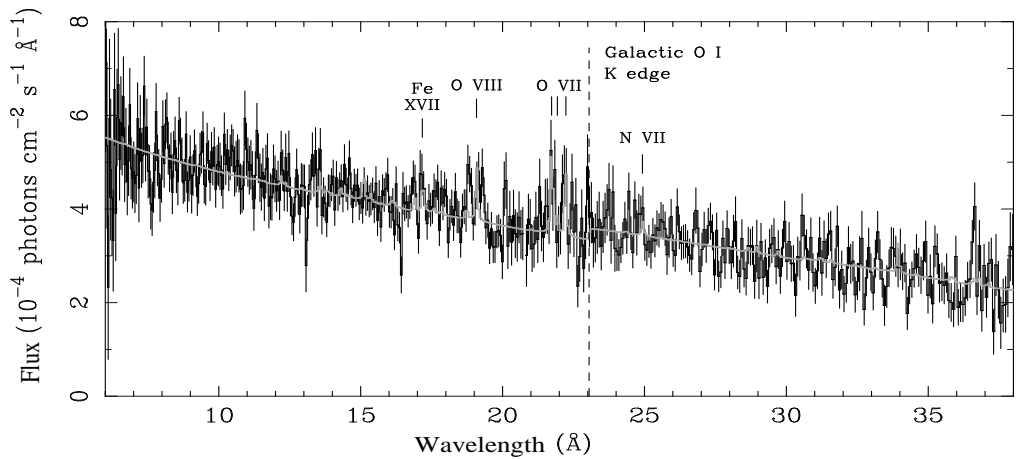}
\caption{The RGS spectrum of NGC~7213 with the best fitting model overplotted in grey. Prominent emission
lines and the O~I K edge from the Galactic ISM are labelled.}
\label{fig:rgsspec}
\end{figure}
\begin{figure}
\centering
\vspace{10.7cm}
\leavevmode
\includegraphics{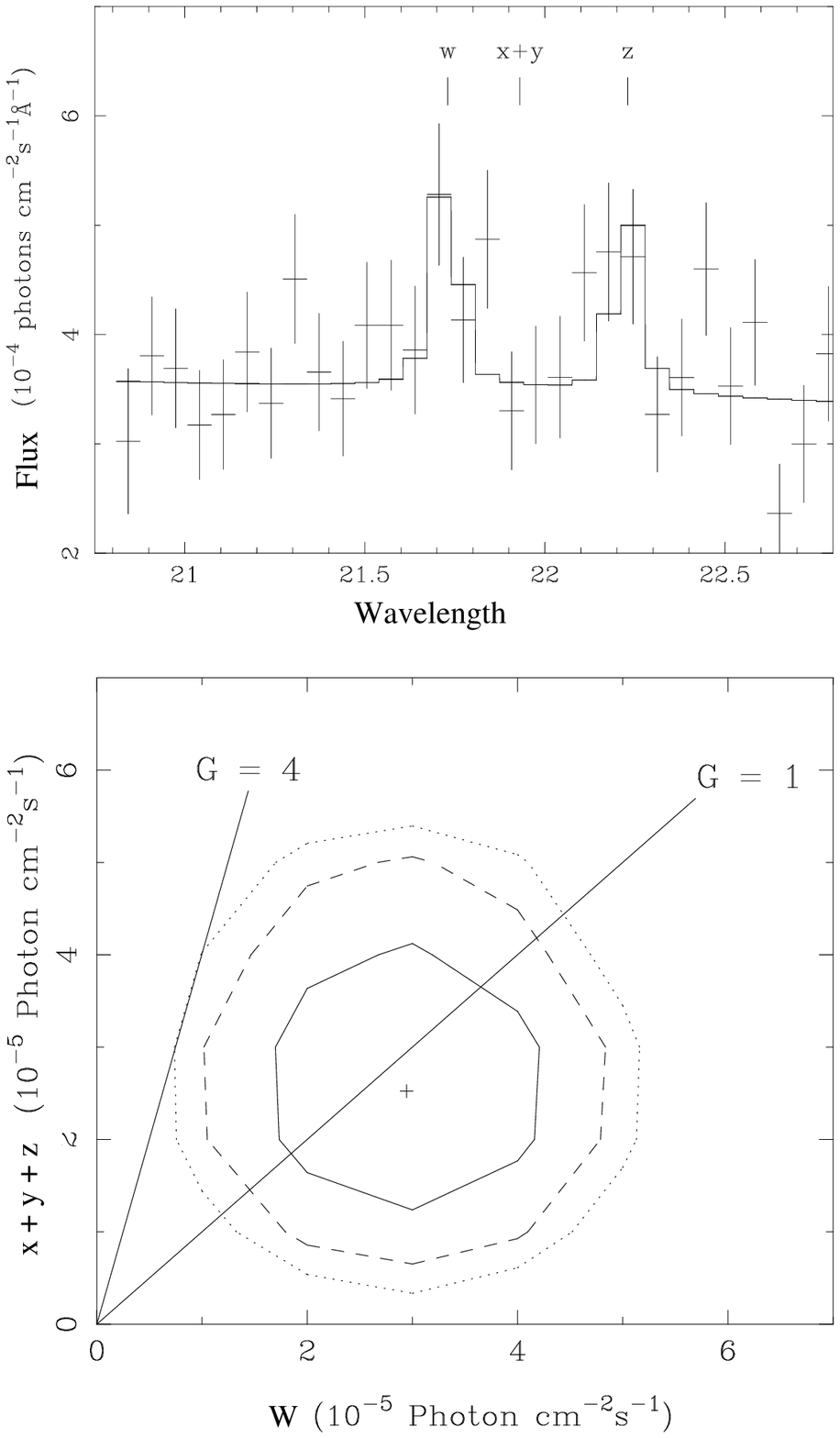}
\caption{Upper panel: close up of the He-like O~VII triplet with best fitting power law plus 3-Gaussian model. Wavelength in \AA. Lower panel: confidence
contours for the strengths of the forbidden and intercombination lines (x+y+z)
and the resonance line (w). The solid, dashed and dotted contour lines
correspond to 68, 90 and 95\% respectively for two interesting parameters. 
$G$$=$1 is expected for a
collisionally ionised plasma; a photoionised plasma should lie to the
left of the $G$$=$4 line.}
\label{fig:ovii}
\end{figure}

RGS:  The RGS spectrum of NGC~7213 is dominated by continuum emission, but emission lines are present, particularly
from O~VII and O~VIII. 
No significant absorption lines or broad absorption features
are observed. The features at 13.08\AA~and 16.43\AA\ are low
signal to noise data points coinciding with
chip-gaps in the first order spectra.
To model the RGS spectrum we began with a power law, which is
rejected at $>$99.5\% confidence. To improve the fit we added a {\small MEKAL}
thermal plasma component, and obtain an acceptable fit with a best fit
plasma temperature of $kT$=$0.18^{+0.02}_{-0.01}$ keV and $\chi^{2}/\nu$=554/496. Addition of a second
thermal plasma component improves the fit only slightly.
The O~VII lines are reproduced well by the 0.18 keV thermal plasma
component, but there appears to be some emission adjacent to O~VIII
Ly$\alpha$ in excess of the model
prediction, perhaps indicating that the higher temperature component is 
 broadened by Doppler motions.
There is no significant blackbody-like soft-excess emission above the power 
law. The model and data are shown in Fig. \ref{fig:rgsspec}.

The `$G$' ratio of the intercombination (x+y) and
forbidden (z) line strengths to the resonance (w) line strength in the He-like triplet of O~VII allows
us to discriminate collisionally ionised and photoionised emission \cite{porquet}. We
obtained the $G$ ratio by fitting the 21-23\AA\ region with a
power law and 3 emission lines (Fig. \ref{fig:ovii}). Collisionally ionised plasmas have $G$$\approx$1, consistent with that observed in NGC~7213
while photoionisation dominated plasmas have $G$$\geq$4, which is 
excluded
at $>$95\% confidence. A photoionised plasma that does not lie along the line
of sight could have $G$$<$4 if the resonance line is enhanced by
photoexcitation \cite{hetrips}. But the $3d-2p$ lines of Fe~XVII at $\sim$15\AA\ should then also be enhanced relative to the
$3s$-$2p$ lines at $\sim$17\AA, as is observed in NGC~1068 \cite{1068}. This is not the case in NGC~7213, 
and so we conclude that the emission lines in the RGS spectrum are predominantly from collisionally ionised gas.

EPIC:  A power law is clearly a poor fit to the 2-10 keV EPIC
data. The presence of reflection has been ruled out in B03, and
these authors conclude that the excess emission is explained with three Fe
emission lines. Combination of the EPIC pn and MOS1 data 
provides better statistics than pn alone. We fit a power law plus 3 Gaussian lines of
fixed narrow width ($\sigma$=1 eV) to the 2-10 keV combined pn-MOS1
spectrum. The best fit ($\chi^{2}$/$\nu$=212/169) has a power law photon
index of $\Gamma$=1.73$\pm$0.01, consistent with that found in the RGS soft X-ray
data.  
The centroid energies of the emission lines in the fit to the combined EPIC
data
are indeed consistent with iron fluorescence in low ionisation material, Fe XXV and Fe XXVI. We find equivalent widths of Fe I (which will include a small contribution from Fe II-XVII), XXV and XXVI K$\alpha$ emission lines of 82$^{+10}_{-13}$, 24$^{+9}_{-11}$ and 24$^{+10}_{-13}$ eV, respectively.
\section{Discussion}
NGC~7213 resembles a typical Seyfert
galaxy, in that its 2-10 keV spectrum is dominated by a $\Gamma$$\sim$1.7 power law and a 6.4 keV 
Fe K$\alpha$ emission line.
Significant emission from Fe~XXV and
Fe~XXVI is also present, which are not normally observed in the classical 
luminous Seyfert galaxies
(eg. NGC~5548, Pounds et al., 2003; NGC~7469,
Blustin et al., 2003), but appear to
dominate the Fe~K$\alpha$ emission in the nearby
LINER M\,81 \cite{M81_2}. 
The Fe~XXV and
Fe~XXVI lines may be produced by photoionisation of
Compton-thin material by the nuclear X-ray source (Bianchi et al., 2004), or may be collisionally ionised like the soft X-ray thermal plasma. 
The soft X-ray emitting gas in Seyfert galaxies is usually found to be photoionised (eg. IRAS 13349+2438, Sako et al., 2001), but unlike typical Seyferts, the soft X-ray spectrum of NGC 7213 is more consistent with the emission from a collisionally ionised plasma.

Many Seyfert galaxies show compelling evidence for an accretion disc 
surrounding the black hole in their X-ray spectra. The main indicators are a soft excess, reflection,
and broad Fe K$\alpha$ line 
emission, all of which originate from the inner parts of the accretion disc. 
None of these indicators are present in the \emph{XMM} spectra of NGC~7213. 
There is no evidence for an optical/UV bump and consequently the AGN bolometric luminosity does not appear to be dominated by emission from an optically-thick,
geometrically-thin accretion disc. From combining our $L_{\rm bol}$ estimate from the \emph{XMM}+archival data SED with the mass estimate of $M_{\rm BH}$=$10^{8.0} \msun$ \cite{NW}, we find that the luminosity of NGC~7213 is low, at approximately 7$\times$10$^{-4}$ $L_{\rm Edd}$.
Therefore it is likely that if there is an accretion disc in NGC~7213, its
inner edge is truncated at a larger radius than is typical in 
Seyfert galaxies. This could be via an ADAF-type flow \cite{ADAF}, or the disc may be in a `low state' (Siemiginowska, Czerny and Kostyunin, 1996).

The lack of reflection (B03) implies the Fe~I~K$\alpha$ line arises in
Compton-thin material. Thus the central region of NGC~7213 appears to be
deficient in the dense, cool material. That LINERs have gas-poor central regions relative to Seyferts has
also been proposed by Ho, Filippenko and Sargent (2003) on the basis of their
optical emission line properties. It appears then that the low luminosities and therefore accretion rates of LINER-AGN are a consequence of a shortage of material in
their central regions. In this case LINERs are just fuel-starved AGN.

If we compare the \emph{XMM} X-ray spectra of NGC~7213 and
the nearest LINER galaxy, M\,81 (Page et al., 2003,2004),
the broad-band X-ray spectra of these two galaxies look
remarkably similar. Whilst the continua are comparable we find
substantial differences in the emission line parameters.
The X-ray spectrum of NGC~7213 is much more Seyfert-like than that of M\,81, owing to the stronger Fe I K$\alpha$ line and weaker soft X-ray lines.
Therefore NGC 7213 appears to bridge the gap
between `normal' Seyfert galaxies and LINER galaxies such as M\,81.

It appears that there is likely a continuous distribution of galaxy nuclei
between the LINERs and `normal' Seyfert nuclei, over which the X-ray spectral 
features characteristic of Seyferts such as the neutral Fe K$\alpha$ line, 
become
successively more prominent, while characteristic LINER features such
as soft X-ray emission lines diminish in significance. 
Accretion rate onto the black hole with respect to the Eddington rate is 
likely to be the overriding factor,
with LINER galaxies accreting at much lower rates than 
Seyfert galaxies (Ho et al., 2003) and containing truncated discs.
In fact, if we look at the observational properties of the Galactic Centre
we see that it may also fit into this continuous distribution since it contains a
low-mass black hole with an extremely low accretion rate: the emission from this region comes predominantly from thermal plasmas
with strong soft X-ray emission \cite{SagA}, and the strongest Fe
K$\alpha$ emission is observed at 6.7 keV \cite{SagAFelines}.

\section{Acknowledgments}
RLCS acknowledges support from a PPARC studentship and an EU RTN fellowship.

\end{article}
\end{document}